\documentclass[a4paper,12pt]{article}

\usepackage{graphics}

\begin{document}

\title{
\bf The solution of the cosmological constant problem from the inhomogeneous equation of state - a hint from modified gravity?}

\author{Hrvoje \v Stefan\v ci\'c\thanks{shrvoje@thphys.irb.hr}
}

\vspace{3 cm}
\date{
\centering
Theoretical Physics Division, Rudjer Bo\v{s}kovi\'{c} Institute, \\
   P.O.Box 180, HR-10002 Zagreb, Croatia}


\maketitle

\abstract{The cosmological constant problem is studied in a two component cosmological model. The universe contains a cosmological constant of an arbitrary size and sign and an additional component with an inhomogeneous equation of state. It is shown that, in a proper parameter regime, the expansion of the universe with a large absolute value of the cosmological constant may asymptotically tend to de Sitter space corresponding to a small effective positive cosmological constant. It is argued that such a behavior can be regarded as a solution of the cosmological constant problem in this model. The mechanism behind the relaxation of the cosmological constant is discussed. A connection with modified gravity theories is discussed and an example of a possible realization of the cosmological constant relaxation in $f(R)$ modified gravity is described.}

\vspace{2cm}

\section{Introduction}

\label{int}

The state of accelerated expansion of the present Universe seems to be better and better 
confirmed by the cosmological observations \cite{SN,WMAP,LSS}. The question of the dynamical mechanism responsible for the accelerated expansion, however, still lacks its definite answer. The last decade has witnessed the arrival of numerous models of the accelerated expansion including dark energy, braneworld models, modified gravity and many others \cite{Rev}. An interesting fact is that when confronted against the observational data, a simple $\Lambda$CDM model, in which the cosmological constant (CC) is the cause of acceleration, fits the data very well. When the conceptual simplicity of the $\Lambda$CDM model is taken into account, it is easy to understand why it is a benchmark model for the analysis of cosmological observations. While it is quite clear why the $\Lambda$CDM model is so appealing from the observational side, its status from the fundamental theoretical perspective is much more problematic. It is by now a notorious fact that the CC value predicted in Quantum Field Theory (QFT) differs from the observed value by an embarrassingly large number of orders of magnitude \cite{Wein}. The problem of explaining the observed value of the cosmological constant is therefore one of the largest challenges in theoretical physics \cite{Wein,Stra,Nob}. This problem is sometimes refered to as {\em the old cosmological constant problem}.  It is further exacerbated by the fact that in all other approaches to the problem of the accelerated expansion of the universe it is assumed that the CC problem is somehow solved. 

Many attempts to solve the CC problem have been made during several last decades \cite{Wein,Stra,Nob}. However, so far none of them has provided a fully satisfactory solution of the CC enigma. The most frequent problem that models of various sorts encounter is the necessity of {\em fine-tuning}. The parameters of the model have to be chosen with extraordinary precision in order for the model to lead to the resolution of the CC problem. Even a very small deviation from these fine-tuned values disrupts the efficiency of the proposed mechanisms.

In this paper we propose a dynamical cosmological model with a specific regime in which it is possible to contemplate the resolution of the CC problem. The proposed model is simplified insofar that it does not contain all the (matter or radiation) components that naturally participate in the evolution of the universe. However, given the difficulty of the CC problem and its resilience to different attempts of solution, it seems preferable to first concentrate on the very mechanism which could produce the observed value of the effective cosmological constant for a universe with values of $\Lambda$ comparable to those predicted in QFT. 

There exists additional problem related to the size of the cosmological constant energy density (or more generally the present dark energy density). Namely, the observational data reveal that the energy density of matter (which at present epoch comes predominantly from nonrelativistic matter) is of the same order of magnitude as the CC (present DE) energy density. Dark energy and nonrelativistic matter scale differently with the expansion of the universe and it is quite remarkable that at present epoch these two energy densities are comparable. This problem is also called {\em the cosmic coincidence problem}. In this paper we are primarily concerned with the problem of the size of the cosmological constant, whereas the cosmic coincidence problem is not addressed.

The principal aim of this paper is to study a two component cosmological model which possesses a cosmological constant of a large absolute value. We investigate conditions under which the asymptotic expansion in this model is of de Sitter type where the asymptotic value of the Hubble function corresponds to a small value of the effective cosmological constant ($H_{asym}^2=\Lambda_{eff}/3$). In the studied models the solution of the cosmological constant is understood as a situation in which the universe ends up in an asymptotic de Sitter regime at large scale factor values characterized by a small cosmological constant (small in a sense that $\Lambda_{eff} \ll |\Lambda|$). The absence of fine-tuning in the model studied in this paper is achieved if the parameters of the model do not have to cancel to many decimal places to lead to the solution of the CC problem. The dynamical process in which the universe in the studied model tends to de Sitter space with a small $\Lambda_{eff}$ is also referred to as the relaxation of the cosmological constant. In the following sections we present a two component model in which the relaxation of the cosmological constant is realized.   

\section{The model set-up}

\label{setup}

We consider the cosmological model described by the FRW metrics containing two components: the cosmological constant with the energy density $\rho_{\Lambda}$ and an additional cosmological component with the energy density $\rho$. Throughout this paper the universe is assumed to be spatially flat $(k=0)$. The Friedmann equation for this cosmological model is
\begin{equation}
\label{eq:H2}
H^2=\frac{8 \pi G}{3} (\rho_{\Lambda}+\rho) \, .
\end{equation}
The equation of state (EOS) of the cosmological constant is standard, $p_{\Lambda}=-\rho_{\Lambda}$. On the other hand, for the equation of state of the second component we take
\begin{equation}
\label{eq:p}
p=w \rho - 3 \zeta_0 H^{\alpha+1} \, ,
\end{equation}
where $p$ is the pressure of the second component and $w$, $\zeta_0$ and $\alpha$ are real parameters. In the considerations given in this paper we limit ourselves to positive values of the parameter $\zeta_0$. Both components satisfy the equation of continuity which for the cosmological constant results in $\rho_{\Lambda}=const$ and for the second components reads as 
\begin{equation}
\label{eq:cont}
d \rho = -3 (\rho+p) \frac{da}{a} \, ,
\end{equation}
where $a$ denotes the scale factor of the universe.

Although the preceding equations of this section describe a simple cosmological model with interesting dynamical regimes, as it will be shown in the following section, it is clear that in Eq. (\ref{eq:p}) lies the nonstandard content of this model which requires physical motivation. This equation of state can be described as an inhomogeneous equation of state in the framework of  Ref. \cite{Odin1} (the same authors consider very similar EOS in a different context). An interesting example where an inhomogeneous dark energy EOS is relevant in the structure formation process is given in \cite{Mota}. Our primary aim is to demonstrate a mechanism of this model which allows a universe with a large $|\Lambda|$ to end up in a de Sitter regime with a small positive $\Lambda_{eff}$. An important issue of a more fundamental basis leading to (\ref{eq:p}) is left for future work. Still, an illustrative example of a possible realization of the relaxation mechanism in terms of $f(R)$ gravity is given in section \ref{grav}. In the remainder of this section we outline two physical frameworks \cite{Odin1} which give motivation for the inhomogeneous equation of state of the type (\ref{eq:p}). It is important to stress that our approach in the study of the model is mainly phenomenological. We primarily focus on the mechanism of the CC relaxation whereas the model studied in this paper is mainly considered as a framework in which the said machanism could be studied. 

The form of Eq. (\ref{eq:p}) is intentionally chosen to emphasize similarity with bulk viscosity. Indeed, the identification $\zeta=\zeta_0 H^{\alpha}$ brings  (\ref{eq:p}) in the form  $p=w \rho - 3 H \zeta$ which is the standard form for the bulk viscosity effects of imperfect cosmological fluid in FRW universe \cite{Weinvisc,Zim}. However, it should be noted that the dependence of $\zeta$ on $H$, which is not a state variable of the fluid, does not correspond to standard bulk viscosity. For the value $\alpha=0$ we recover the bulk viscous imperfect fluid with the constant coefficient $\zeta$. Therefore, we are motivated by the bulk viscosity, we consider its generalization and then proceed phenomenologically.  Here it would be preferable to call the inhomogeneous term the nonlinear (bulk) viscosity \cite{Gron}. Therefore, a possible identification of the second component might be as an imperfect cosmological fluid with nonlinear viscosity having a power-law dependence on the Hubble parameter $H$.

An alternative view is to interpret (\ref{eq:p}) as an inhomogeneous equation of state coming from modified gravity or braneworld models (see Appendix in \cite{Odin1}). The considerations given in section \ref{grav} lend support to this interpretation.  

In the following section we focus on the dynamical regimes of the system of equations (\ref{eq:H2}), (\ref{eq:p}) and (\ref{eq:cont}) and the related phenomena including the possibility of the relaxation of a large cosmological constant (large in terms if its absolute value). 




\section{The model dynamics}

\label{dyn}

The model defined in the preceding section is now analyzed in detail. Combining (\ref{eq:H2}) and (\ref{eq:p}) with (\ref{eq:cont}) we obtain
a dynamical equation for the Hubble parameter
\begin{equation}
\label{eq:dynH}
d H^2 + 3(1+w)\frac{da}{a} \left( H^2 - \frac{8 \pi G \rho_{\Lambda}}{3} - \frac{8 \pi G \zeta_0}{1+w} (H^2)^{(\alpha+1)/2} \right) =0\, .
\end{equation}
The analysis of this equation is further simplified by the introduction of the following notation: 
\begin{equation}
\label{eq:notation}
h=(H/H_X)^2, \;\; s=a/a_X, \;\; \lambda=8 \pi G \rho_{\Lambda}/3 H_X^2, \;\; \xi=8 \pi G \zeta_0 H_X^{\alpha-1}/(1+w)\, . 
\end{equation}
Here $H_X$ denotes the value of the Hubble parameter at, in principle arbitrary, value of the scale factor $a_X$. Let us further stress that although we assume $\zeta_0>0$, the parameter $\xi$ may take values of both signs if we also allow the values $w<-1$.  Applying the described change of notation, Eq. (\ref{eq:dynH}) acquires the form
\begin{equation}
\label{eq:dynH2}
s \frac{d h}{d s} + 3 (1+w) (h - \lambda - \xi h^{\frac{\alpha+1}{2}})=0 \, ,
\end{equation}
with the initial condition $h(1)=1$.

The inspection of Eq. (\ref{eq:dynH2}) reveals that the value of the parameter $\alpha$ may significantly influence the type of dynamics of the Hubble parameter. The values $\alpha=-1$ and $\alpha=1$ are specific points at which we expect the change of dynamical behavior. Therefore we analyze five characteristic intervals/points for $\alpha$: $(-\infty,-1),-1,(-1,1),1$ and $(1,\infty)$. For each of the intervals/points we make an analytical treatment at one value of the parameter $\alpha$ and, when necessary, support it with numerical calculations.

\subsection{$\alpha < -1$: the relaxation mechanism for a large cosmological constant}

\label{mainres}

The analysis of the interval $\alpha < -1$ reveals a dynamical mechanism for the relaxation of a large cosmological constant to a much smaller effective CC value. The contents of this subsection comprise the main results of the present paper. We start the analysis of the Hubble dynamics with an analytical treatment for a representative value $\alpha=-3$. Eq. (\ref{eq:dynH2}) now becomes
\begin{equation}
\label{eq:alpha-3}
\frac{h \, d h}{h^2-\lambda h - \xi} = -3 (1+w) \frac{d s}{s} \, .
\end{equation}
The integration of the left hand side of Eq. (\ref{eq:alpha-3})  is determined by the zeros of its denominator which we denote by $h_{*1,2}$, i.e. $h^2-\lambda h - \xi=(h-h_{*1})(h-h_{*2})$. Their specific values are
\begin{equation}
\label{eq:hstar1}
h_{*1}= \frac{1}{2} \left(\lambda+\sqrt{\lambda^2+4 \xi}\right)\, , \\
\end{equation}
\begin{equation}
\label{eq:hstar2}
h_{*2}= \frac{1}{2} \left(\lambda-\sqrt{\lambda^2+4 \xi}\right) \, . 
\end{equation}
Generally we have $h_{*1}-h_{*2}>0$. 
The solution of (\ref{eq:alpha-3}) is of the form
\begin{equation}
\label{eq:solrelax}
\left( \frac{h-h_{*1}}{1-h_{*1}} \right)^{A_{1}} \left( \frac{h-h_{*2}}{1-h_{*2}} \right)^{A_{2}} = s^{-3(1+w)} \, ,
\end{equation}
where $A_1=h_{*1}/(h_{*1}-h_{*2})$ and  $A_2=-h_{*2}/(h_{*1}-h_{*2})$. Let us further study separately cases of positive and negative values of the parameter $\xi$. 

For $w>-1$ the value of the parameter $\xi$ is positive. In this case we have $h_{*1}>0$ and $h_{*2}<0$ which leads to $A_1>0$ and $A_2>0$. Eq. (\ref{eq:solrelax}) readily provides information on the asymptotic behavior of the parameter $h$. Namely, for small values of the scale factor, $s \rightarrow 0$ the function $h$ diverges
\begin{equation}  
\label{eq:lamnegs0}
h \sim (1-h_{*1})^{A_1}(1-h_{*2})^{A_2} s^{-3(1+w)} \, .
\end{equation}
For large values of the scale factor the function $h$ tends to a constant value
\begin{equation}
\label{eq:lamnegslarge}
\lim_{s \rightarrow \infty} h = h_{*1} \, .
\end{equation}

As stated in (\ref{eq:lamnegslarge}), the parameter $h$ (equivalently the Hubble parameter squared $H^2$) asimptotically tends to a constant value at large values of the scale factor. Let us consider the case when $\lambda$ is negative and very large in absolute value compared to $\sqrt{\xi}$, or, more precisely, $\lambda^2 \gg 4 \xi$. The square root in the expression (\ref{eq:hstar1}) can now be expanded and we obtain
\begin{equation}
\label{eq:happrox1}
h_{*1} \simeq \frac{\xi}{|\lambda|} \, ,
\end{equation}
or, equivalently,
\begin{equation}
\label{eq:Hstareq}
H_{*1}^2=\frac{24 \pi G \zeta_0}{(1+w) |\Lambda|} \equiv \frac{3 \zeta_0}{(1+w) |\rho_{\Lambda}|}\, .
\end{equation}

For $\rho_{\Lambda}$ very large in absolute value and negative, the universe asymptotically tends to a small value of $H^2$ which can be interpreted in a straightforward manner as a small value of the effective cosmological constant. It is sufficient that $\Lambda$ is negative and large in absolute value and that $\xi$ is sufficiently small. The model under study in this paper, therefore, provides a dynamical mechanism for the relaxation of a negative $\Lambda$ with a very large absolute value. The dynamical mechanism does not incorporate fine-tuning of model parameters and it can be considered as a solution of the cosmological constant problem in this model for a negative large $\Lambda$. 

Although the choice $\alpha=-3$ gives an analytically tractable example of the relaxation of a negative $\Lambda$, we further support the findings of the preceding paragraph with numerical solutions of Eq. (\ref{eq:dynH2}) for other values in the interval $\alpha < -1$.   
 
\begin{figure}
\centerline{\resizebox{0.6\textwidth}{!}{\includegraphics{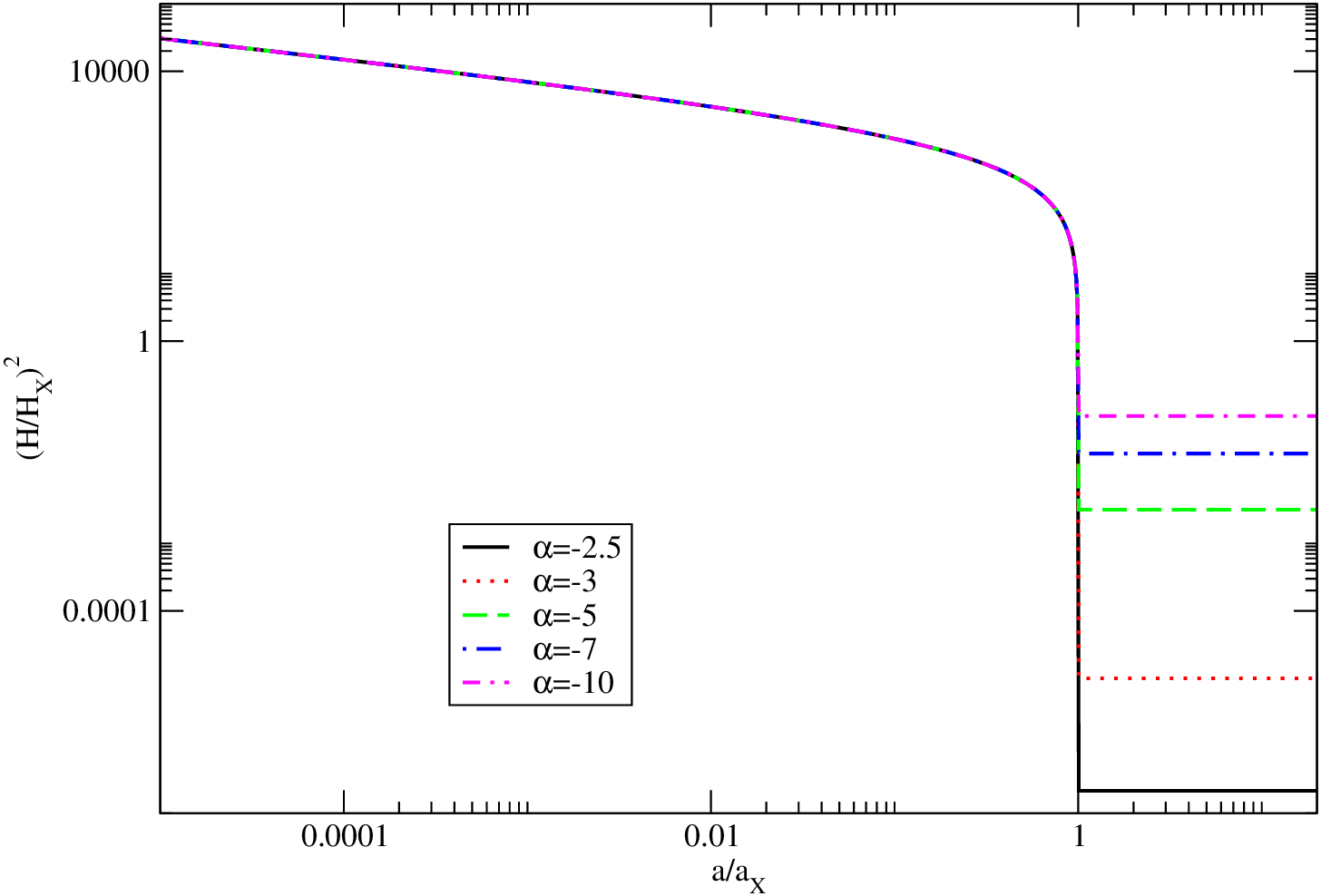}}}
\caption{\label{fig:lnega} The evolution of $h=H^2/H_X^2$ as a function of the scale factor for different values of the exponent $\alpha$. The value of $\alpha$ strongly influences the asymptotic value of $h$ at large $a$, whereas the behavior at small $a$ is not affected by $\alpha$. The values of the parameters used are $\lambda=-1000$, $\xi=0.01$ and $w=-0.9$.}
\end{figure}

In Fig. \ref{fig:lnega} we present the evolution of the variable $h=H^2/H_X^2$ as a function of the scale factor for different values of the exponent $\alpha$. The most striking feature of the dynamics of $h$ for all studied values of $\alpha$, is the abrupt transition between two asymptotic regimes. This feature could not be properly addressed from the study of the asymptotic regimes alone, but it requires a numerical treatment to be fully appreciated. The value of exponent $\alpha$ does not affect the asymptotic evolution at small $a$, but it crucially affects the large $a$ asymptotic behavior.   

\begin{figure}
\centerline{\resizebox{0.6\textwidth}{!}{\includegraphics{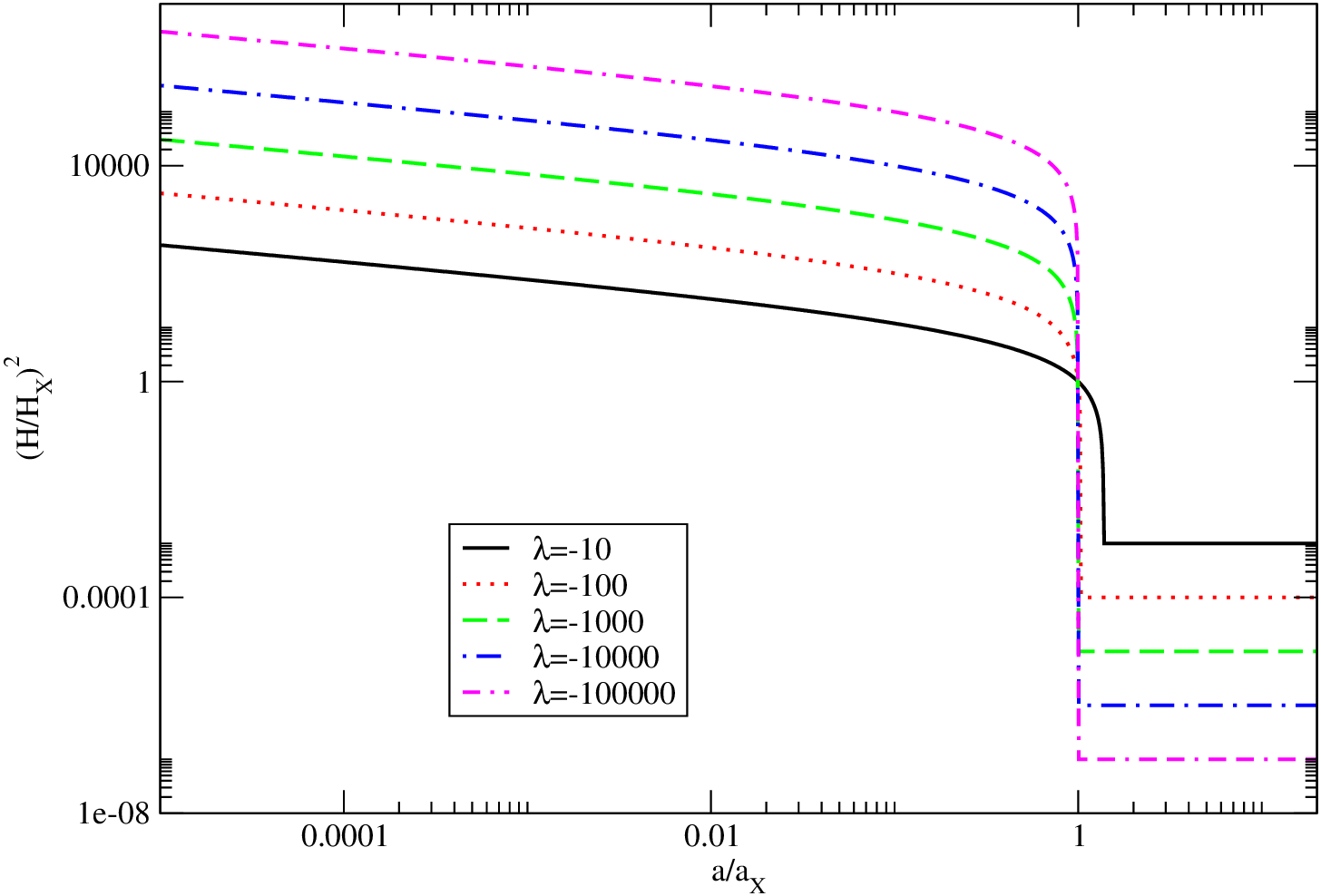}}}
\caption{\label{fig:lnegl} The dynamics of $h=H^2/H_X^2$ as a function of the scale factor for different values of the CC parameter $\lambda$. The value of $\lambda$ affects the behavior at small values of $a$, large values of $a$ and the onset of the transition between two regimes. The values of the parameters used are $\alpha=-3$, $\xi=0.01$ and $w=-0.9$.
}
\end{figure}

Figure \ref{fig:lnegl} depicts the dependence of the dynamics of $h$ on the CC parameter $\lambda$. It is evident that the value of $\lambda$ affects both asymptotic regimes (at large and small scale factor values) as well as the onset of the abrupt transition between two regimes.

\begin{figure}
\centerline{\resizebox{0.6\textwidth}{!}{\includegraphics{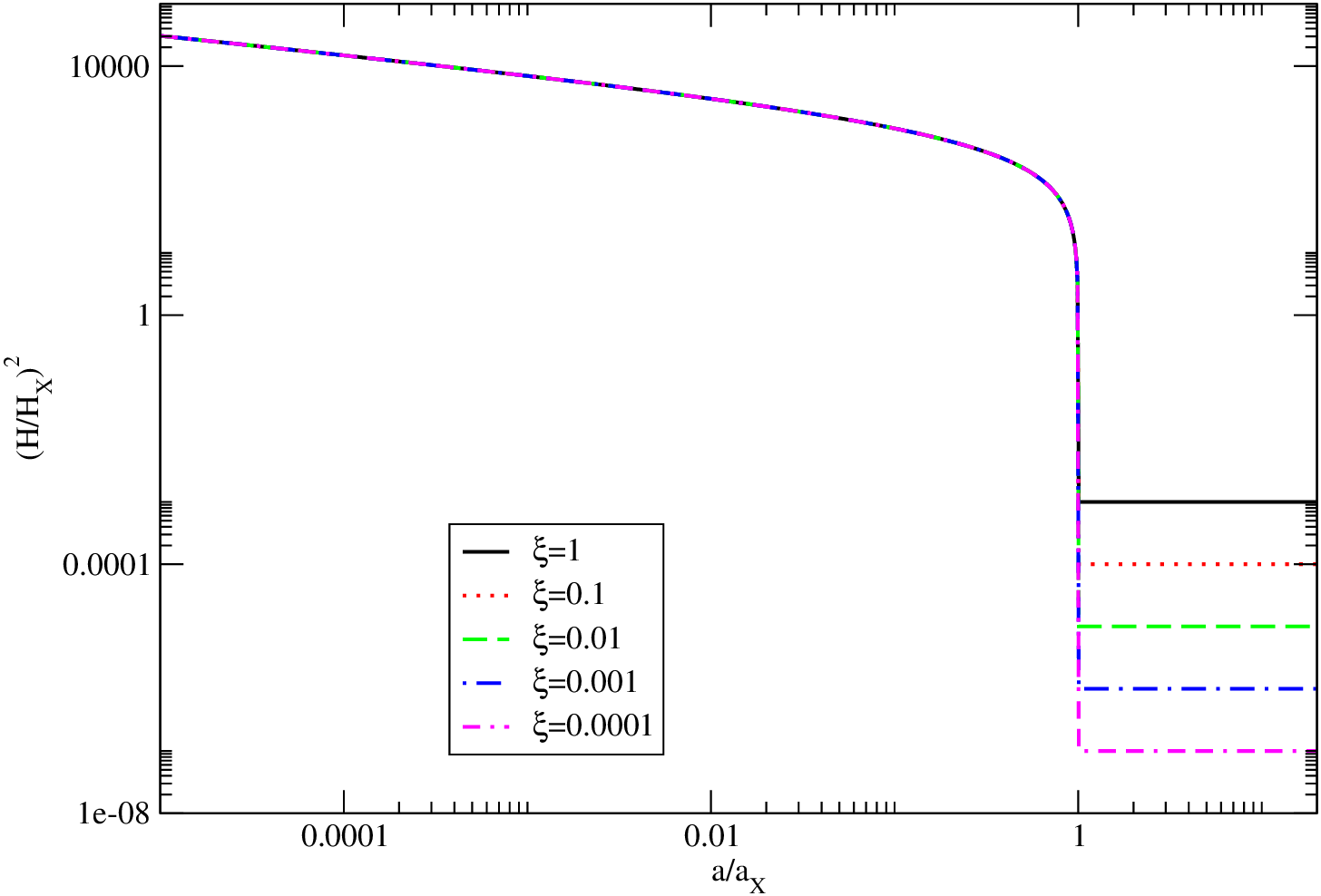}}}
\caption{\label{fig:lnegk} The dependence of $h$ as a function of the scale factor on the parameter $\xi$. It is evident that the behavior at small values of $a$ does not depend on $\xi$, but the asymptotic dynamics at large $a$ is strongly affected by the value of the parameter $\xi$. The values of the parameters used are $\lambda=-1000$, $\alpha=-3$ and $w=-0.9$. }
\end{figure}

In Fig. \ref{fig:lnegk} we study the dependence of the behavior of $h$ on the parameter $\xi$. From the figure it is clear that the dynamics of $h$ at small values of scale factor does not depend on $\xi$, but at large values of $a$ the asymptotic value of $h$ is strongly influenced by the value of $\xi$. 

\begin{figure}
\centerline{\resizebox{0.6\textwidth}{!}{\includegraphics{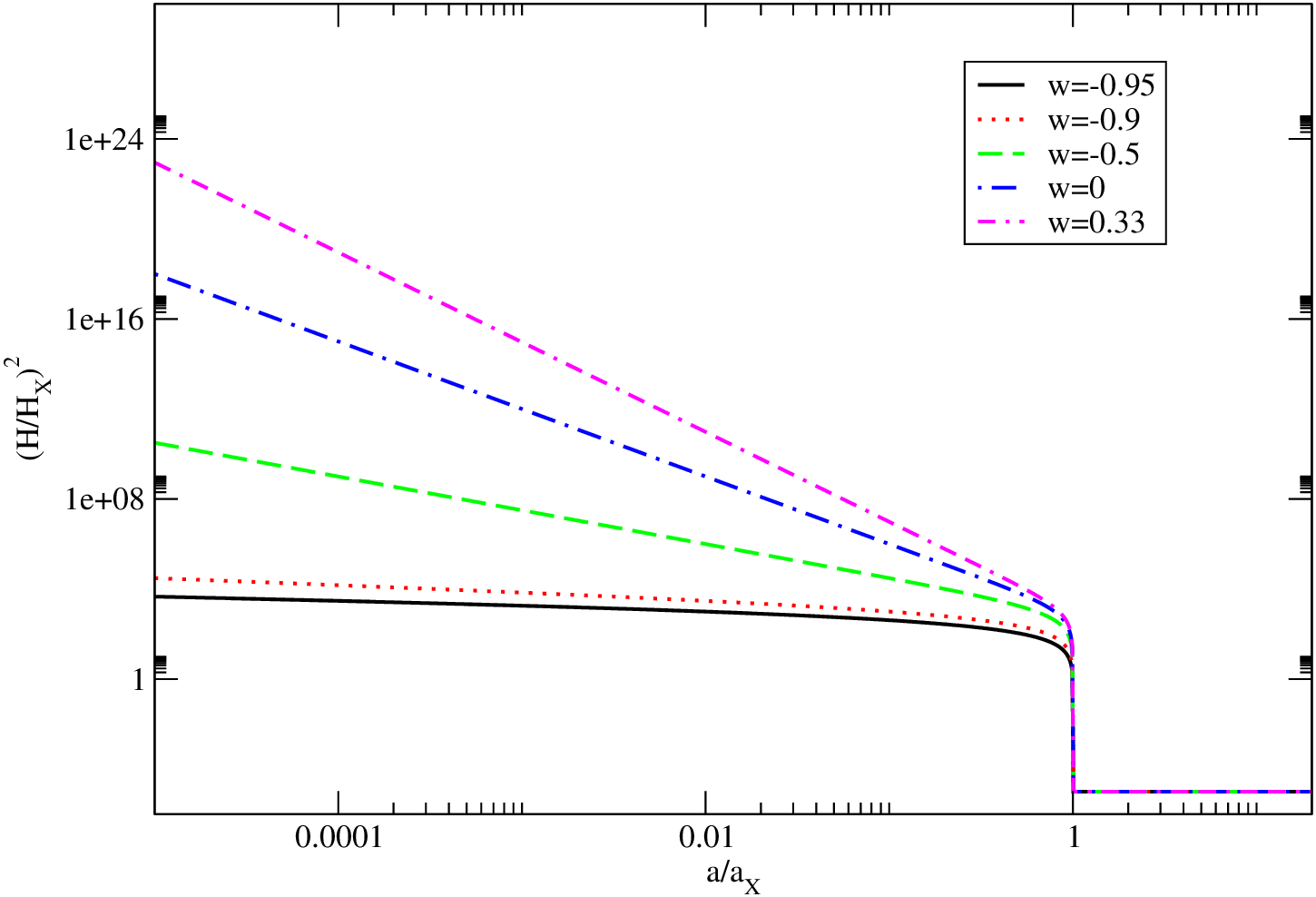}}}
\caption{\label{fig:lnegw}  The evolution of $h$ as a function of scale factor for different values of the parameter $w$. The dynamics at small $a$ is clearly affected by the value of $w$ and the dynamics at large values of $a$ is insensitive to parameter $w$. The values of the parameters used are $\lambda=-1000$, $\alpha=-3$ and $\xi=0.01$. 
}
\end{figure}

Finally, in Fig. \ref{fig:lnegw} we present the dynamics of $h$ as a function of the scale factor $a$ for various values of the parameter $w$. The plots in the figure reveal that the behavior at small $a$ is strongly affected by $w$, whereas the asymptotic behavior at large scale factor values does not depend on $w$. 

Next we return to our analytically tractable case of $\alpha = -3$, but this time we consider a positive value of the cosmological constant $\lambda$. 
We also choose $w<-1$ so that the parameter $\xi$ becomes negative. In this setting we have $h_{*1}>0$ and $h_{*2}>0$ which results in $A_1>0$ and $A_2<0$. The asymptotic behavior of the Hubble parameter now obeys the following laws:
\begin{equation}
\label{eq:lampoys0}
\lim_{s \rightarrow 0} h = h_{*1} \, 
\end{equation}
and
\begin{equation}
\label{eq:lampozslarge}
\lim_{s \rightarrow \infty} h = h_{*2} \, .
\end{equation}
We see that for large values of the scale factor the scaled Hubble parameter tends to a constant value $h_{*2}$. For a very large values of $\lambda$ (such that $\lambda^2 \gg -4 \xi$) the universe asymptotically acquires a small value 
\begin{equation}
\label{eq:happrox2}
h_{*2} \simeq -\frac{\xi}{\lambda} \, ,
\end{equation}
or, equivalently,
\begin{equation}
\label{eq:Hstareq2}
H_{*2}^2=-\frac{24 \pi G \zeta_0}{(1+w) \Lambda} \equiv -\frac{3 \zeta_0}{(1+w) \rho_{\Lambda}}\, .
\end{equation}
For a very large value $\rho_{\Lambda}$ and a small value of $|\xi|$ the asymptotic value (\ref{eq:Hstareq2}) is very small. This small value of $H^2$ can be directly interpreted as a small effective positive cosmological constant. As for the case of negative $\lambda$, we have at hand a dynamical mechanism of the cosmological constant relaxation which does not incorporate fine-tuning of model parameters. Therefore, within the model studied in this paper, we have presented a solution of the CC problem for a positive cosmological constant. It is important to notice that $\rho$ in this case must be negative. This fact implies that the component with an energy density $\rho$ is an effective description of some other dynamical mechanism, possibly modified gravity.

As for the case of negative CC, we further support the analytical treatment for $\alpha=-3$ with numerical analyses for other values in the interval $\alpha <-1$ and other model parameters.   

\begin{figure}
\centerline{\resizebox{0.6\textwidth}{!}{\includegraphics{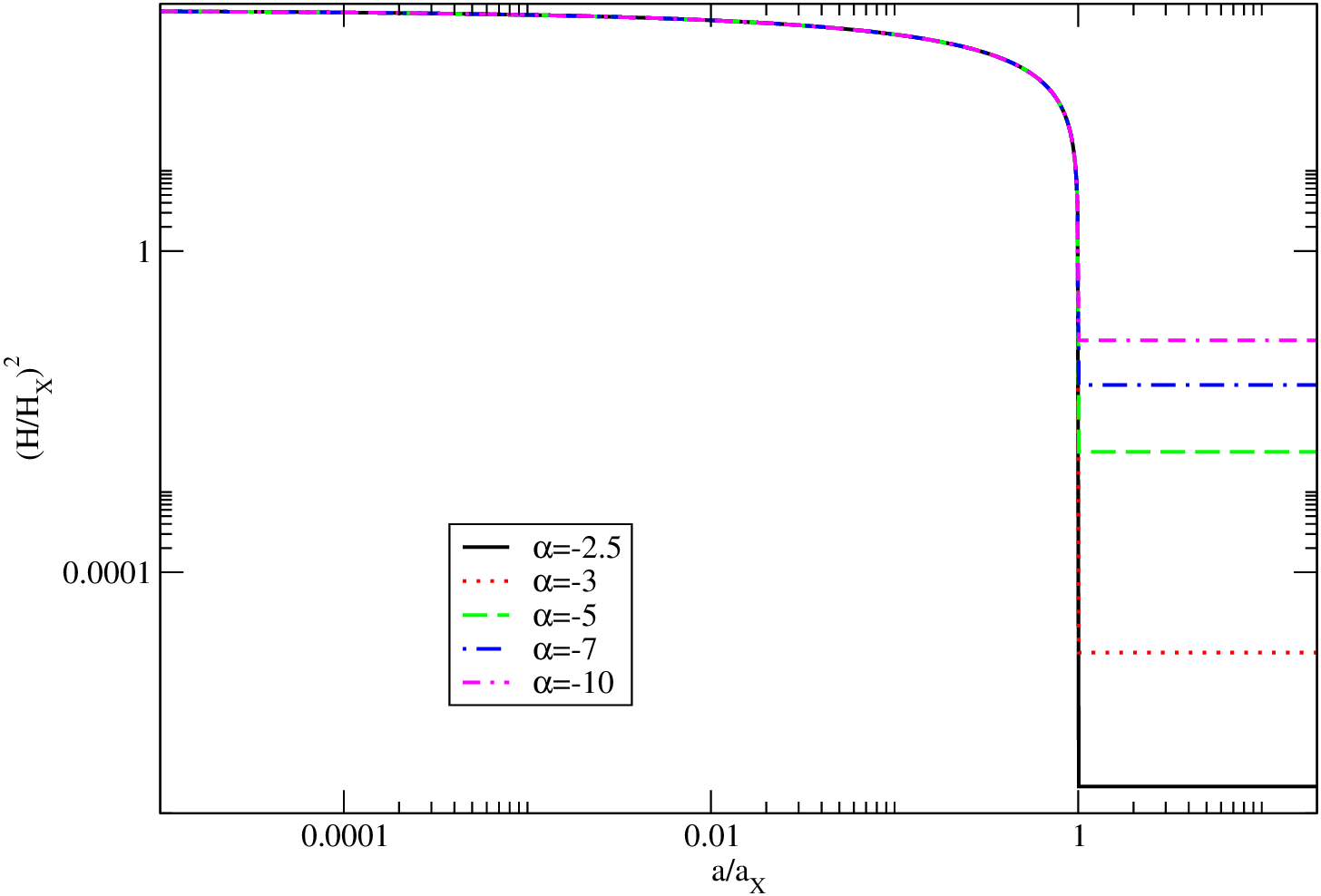}}}
\caption{\label{fig:lposa}  The dependence of $h$ on scale factor for different values of the exponent $\alpha$. The graphs reveal that the behavior of $h$ at small values of $a$ is insensitive to the value of $\alpha$, whereas at large values of $a$ there is a strong dependence of the asymptotic value of $h$ on $\alpha$. The values of the other model parameters used are $\lambda=1000$, $\xi=-0.01$ and $w=-1.1$. 
}
\end{figure}

In Fig. \ref{fig:lposa} the dependence of $h$ on the scale factor for different values of the exponent $\alpha$ is depicted. The behavior of $h$ for small values of the scale factor shows no dependence on the exponent $\alpha$. At large values of the scale factor the asymptotic value of $h$ is strongly influenced by the value of $\alpha$. As in the case with the negative CC with a large absolute value, here the transition between dynamical regimes at small and large values of $a$ is abrupt. Another significant difference compared to the case  of negative CC is that both at small and large values of the scale factor the expansion is of de Sitter type (for negative CC we have a de Sitter-like expansion only at large values of the scale factor).  

\begin{figure}
\centerline{\resizebox{0.6\textwidth}{!}{\includegraphics{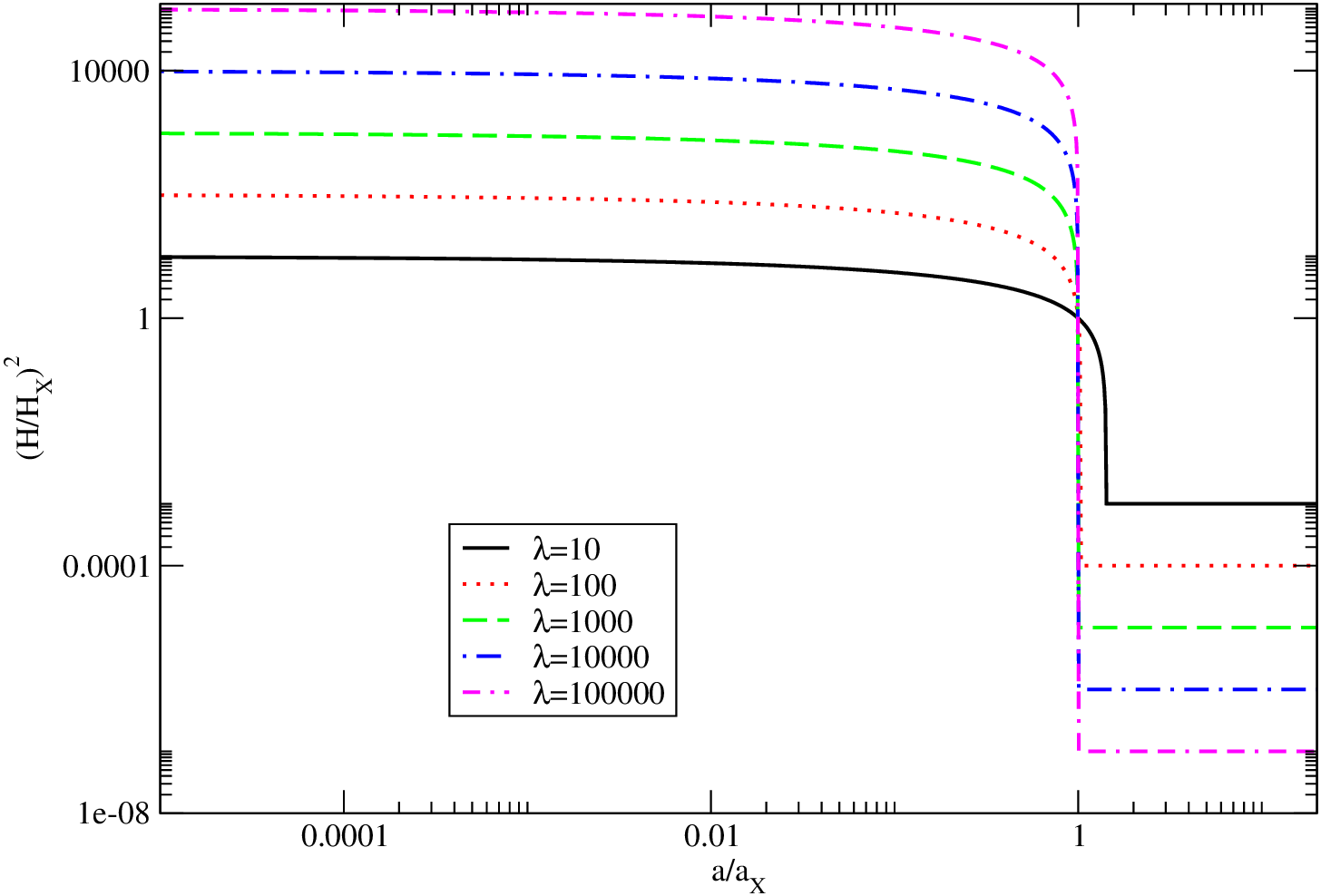}}}
\caption{\label{fig:lposl}  The dynamics of $h$ as a function of the scale factor for various values of the parameter $\lambda$.
The CC parameter $\lambda$ affects the behavior at small and large values of the scale factor as well as the onset of the abrupt transition between two asymptotic regimes. The values of the other parameters used are $\alpha=-3$, $\xi=-0.01$ and $w=-1.1$. 
}
\end{figure}

The dependence of $h$ on the scale factor for various values of the CC parameter $\lambda$ is given in Fig. \ref{fig:lposl}.
The value of $\lambda$ affects the dynamics at small and large values of $a$ and the onset of the abrupt transition between two regimes.

\begin{figure}
\centerline{\resizebox{0.6\textwidth}{!}{\includegraphics{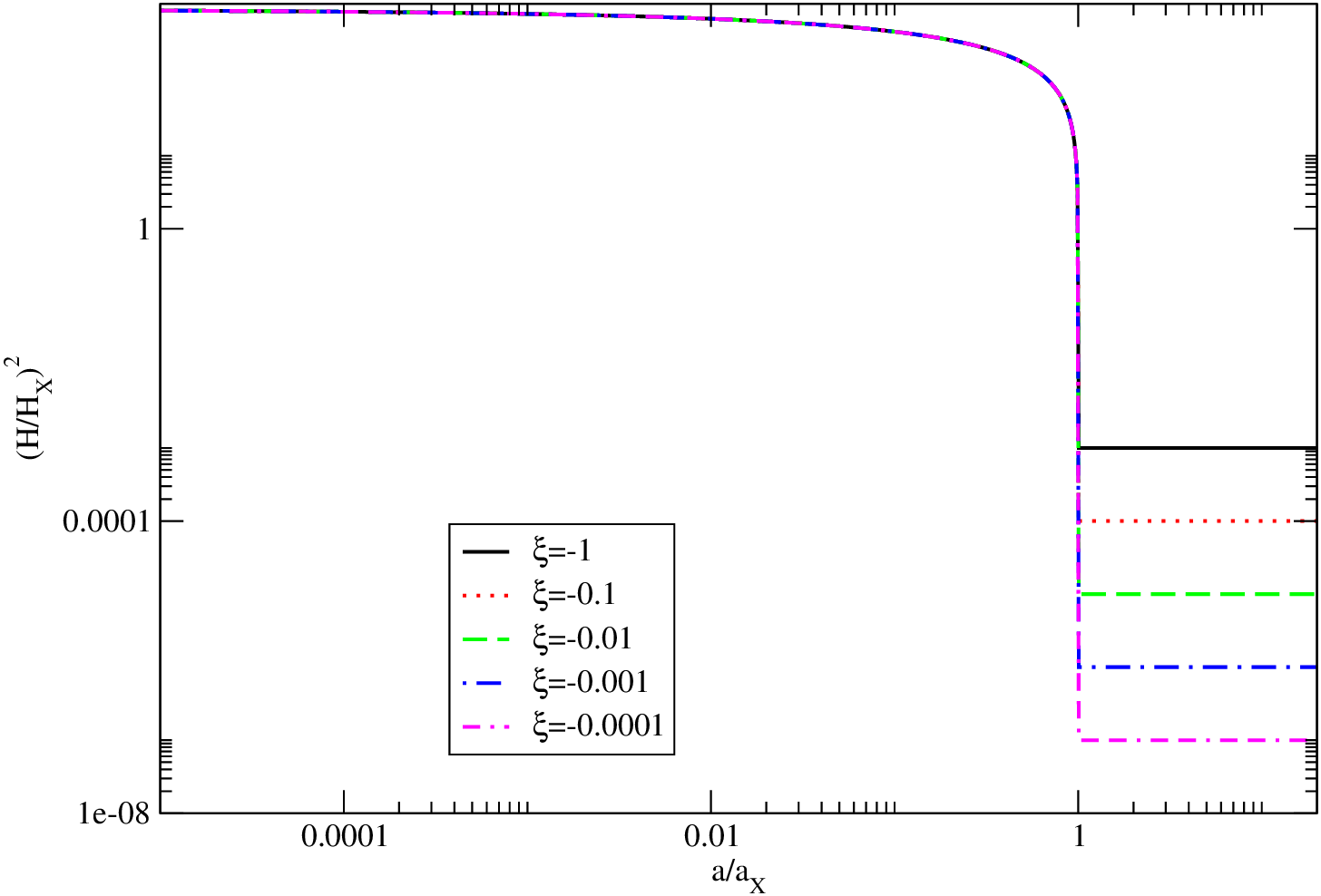}}}
\caption{\label{fig:lposk} The dependence of $h$ on the scale factor for various values of the parameter $\xi$. The dynamics at small $a$ is not influenced by $\xi$, but at large $a$ the asymptotic behavior is sensitive to the value of $\xi$. The values of the parameters used in this figure are $\alpha=-3$, $\lambda=1000$ and $w=-1.1$. 
}
\end{figure}

The sensitivity of the dynamics of $h$ as a function of $a$ on the parameter $\xi$ is presented in Fig. \ref{fig:lposk}.
The dynamics at small values of $a$ is not affected by the value of $\xi$, but at large values of the scale factor the asymptotic value of $h$ depends strongly on $\xi$. 

\begin{figure}
\centerline{\resizebox{0.6\textwidth}{!}{\includegraphics{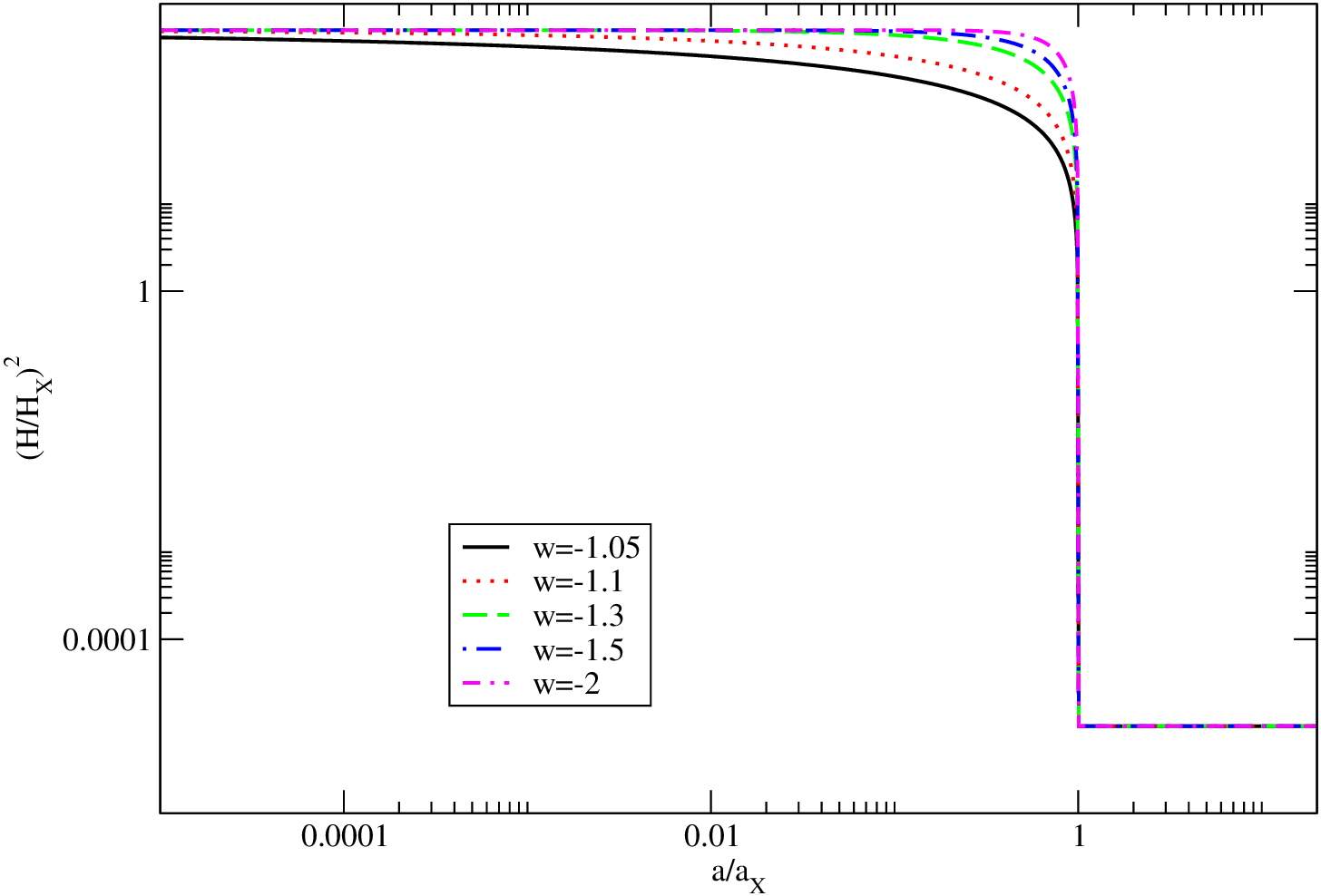}}}
\caption{\label{fig:lposw} The dependence of $h$ as a function of $a$ for different values of the parameter $w$. This parameter affects the form of transition between asymptotic values of $h$ which are insensitive to $w$. The values of the parameters used  are $\alpha=-3$, $\lambda=1000$ and $\xi=-0.01$.  
}
\end{figure}

Finally, in Fig. \ref{fig:lposw} the dependence of $h$ on the scale factor for several values of $w$ is depicted. The plots in the figure reveal that the value of $w$ does not affect the asymptotic values of $h$ at small and large values of the scale factor, but they do affect the transition between these asymptotic values. For all studied values of $w$ this transition remains quite abrupt. 

\subsection{$\alpha = -1$}

The solution of (\ref{eq:dynH2}) for $\alpha=-1$ can be expressed in a familiar form 
\begin{equation}
\label{eq:alpha-1}
h= \lambda + \xi + (1-\lambda-\xi) s^{-3(1+w)} \, .
\end{equation}
This expression describes the universe with the cosmological constant energy density $ \sim \frac{3}{8 \pi G} (\lambda+\xi)$ and the matter component with the scaling $\sim a^{-3(1+w)}$. Clearly, for $w=0$ our model fully mimics the $\Lambda$CDM model. If we wish to have a small effective value of the cosmological constant $\sim \lambda + \xi$, for a large value of $\lambda$ we need to have a cancellation of $\lambda$ and $\xi$ which necessarily introduces fine-tuning. Therefore, for $\alpha=-1$ we do not have an efficient mechanism for the relaxation of the cosmological constant.   

\subsection{$-1 < \alpha < 1$}

In this interval for the value of $\alpha$ we consider the case $\alpha=0$. The Eq. (\ref{eq:dynH2}) now reads 
\begin{equation}
\label{eq:alpha0}
\frac{ y \, dy}{y^2-\xi y - \lambda} = -\frac{3}{2} (1+w) \frac{ds}{s} \, ,
\end{equation}
where $y=h^{1/2}$. The solution of this equation can be cast in the form
\begin{equation}
\label{eq:solalpha0}
\left( \frac{y-y_{*1}}{1-y_{*1}} \right)^{B_1}  \left( \frac{y-y_{*2}}{1-y_{*2}} \right)^{B_2} = s^{-\frac{3}{2}(1+w)} \, ,
\end{equation}
where
\begin{equation}
\label{eq:y1}
y_{*1}=\frac{1}{2} \left(\xi + \sqrt{\xi^2+4 \lambda} \right) \, ,
\end{equation}
\begin{equation}
\label{eq:y2}
y_{*2}=\frac{1}{2} \left(\xi - \sqrt{\xi^2+4 \lambda} \right) \, ,
\end{equation} 
with $B_1=y_{*1}/(y_{*1}-y_{*2})$ and $B_2=-y_{*2}/(y_{*1}-y_{*2})$.
We again study two interesting cases
\begin{enumerate}
\item We consider $w<-1$ where we have $\xi<0$. For the value of the cosmological constant $\lambda$ let us take a positive value which is sufficiently small, so that $\xi^2 \gg 4 \lambda$. Now we have $y_{*1}>0$, $y_{*2}<0$ with $y_{*1}-y_{*2}>0$ and $B_{1,2}>0$.
The asymptotic behavior of the system is given by $\lim_{s \rightarrow 0} y = y_{*1}$ and $\lim_{s \rightarrow \infty} y =  \infty$ with $ y \sim (1-y_{*1})^{B_1} (1-y_{*2})^{B_2} s^{-3(1+w)/2}$. At large values of the scale factor we do not have a de Sitter regime although there is one at small values of $a$. 
\item Another interesting regime is obtained for $w>-1$ which implies $\xi>0$. We study the case of a negative $\lambda$ which is again sufficiently small in absolute value i.e. $\xi^2 \gg 4 |\lambda|$. Then we have $y_{*1}>0$, $y_{*2}>0$ with $y_{*1}-y_{*2}>0$ and $B_{1}>0$, $B_{2}<0$. The asymptotic behavior of the system is $\lim_{s \rightarrow 0} y = y_{*2}$ and $\lim_{s \rightarrow \infty} y = y_{*1}$. From (\ref{eq:y1}) it is easy to see that $y_{*2} \simeq -\lambda/\xi$ and $y_{*1} \simeq \xi$. For a large value $|\lambda|$, $\xi^2$ has to be even larger and therefore at large values of the scale factor we have $H^2 \sim y_{*2}^2 \sim \xi^2$ what does not correspond to a small value of effective CC.

\end{enumerate}

\subsection{$\alpha = 1$}

For $\alpha=1$  and $\xi \neq 1$ the solution of Eq. (\ref{eq:dynH2}) acquires the form
\begin{equation}
\label{eq:alpha1}
h= \frac{\lambda}{1-\xi} + \left( 1-\frac{\lambda}{1-\xi} \right) s^{-3(1+w)(1-\xi)} \, .
\end{equation} 
There are two cases of interest for the CC problem:
\begin{enumerate}
\item For $w<-1$ we have $\xi < 0$. Let us further consider the case $\lambda>0$ with $\lambda/(1-\xi) \ll 1$. Under these conditions the asymptotic behavior of $h$ is the following: at small values of the scale factor we have $\lim_{s \rightarrow 0} h = \lambda/(1-\xi)$, whereas for large values of the scale factor we have $\lim_{s \rightarrow \infty} h = \infty$ with $h \sim (1-\lambda/(1-\xi)) s^{-3(1+w)(1-\xi)}$. In this case there is no asymptotic de Sitter solution for large values of $a$. 
\item If $w>-1$ then $\xi > 0$. Let us take $1-\xi<0$ with $\xi \ll 1$ and $\lambda/(1-\xi) \ll 1$. Then we again have $\lim_{s \rightarrow 0} h = \lambda/(1-\xi)$ and  $\lim_{s \rightarrow \infty} h = \infty$ with $h \sim (1-\lambda/(1-\xi)) s^{-3(1+w)(1-\xi)}$. Again there is no realization of the scenario of interest. 
\end{enumerate} 

For $\alpha=1$ and $\xi=1$ the solution for the dynamics of $h$ becomes
\begin{equation}
\label{eq:alpha1xi1}
h =1+3 \lambda (1+w) \ln s \, .
\end{equation}
This solution does not asymptotically lead to de Sitter space at large scale factor values.

\subsection{$\alpha > 1$}  

As a representative and analytically tractable case in this interval of the exponent $\alpha$ we study the dynamics of $h$ for the value $\alpha=3$. The dynamical equation (\ref{eq:dynH2}) can now be written
\begin{equation}
\label{eq:dynalpha3}
\frac{d \, h}{h^2 - \frac{h}{\xi} + \frac{\lambda}{\xi}}=3(1+w)\xi \frac{ds}{s} \, .
\end{equation}
Introducing the notation $h^2-h/\xi+\lambda/\xi=(h-h_{*1})(h-h_{*2})$ we have
\begin{equation}
\label{eq:zero1alpha3}
h_{*1}=\frac{1}{2}\left( \frac{1}{\xi} + \sqrt{\frac{1}{\xi^2}-\frac{4 \lambda}{\xi}} \right) \, ,
\end{equation}
\begin{equation}
\label{eq:zero2alpha3}
h_{*1}=\frac{1}{2}\left( \frac{1}{\xi} - \sqrt{\frac{1}{\xi^2}-\frac{4 \lambda}{\xi}} \right) \, ,
\end{equation}
which leads to the following solution: 
\begin{equation}
\label{eq:solalpha3}
\frac{h-h_{*1}}{h-h_{*2}} = \frac{1-h_{*1}}{1-h_{*2}} s^{3 (1+w) \xi (h_{*1}-h_{*2})} \, .
\end{equation}
Let us now restrict ourselves to the parameters satisfying $1-4 \xi \lambda > 0$ as a prerequisite for having real asymptotic values for $h$. Again we consider two cases:
\begin{enumerate}
\item For $w>-1$ which implies $\xi>0$ the condition of reality of asymptotic values of $h$ leads to the requirement $\lambda< 1/(4 \xi)$. We have $h_{*1}-h_{*2}>0$, $h_{*1}>0$ whereas $h_{*2} >0$ for $0<\lambda<1/(4 \xi)$ and $h_{*2}<0$ for $\lambda<0$.
For $\lambda>0$ we have $\lim_{s \rightarrow 0} h = h_{*1}$ and $\lim_{s \rightarrow \infty} h = h_{*2}$. For $\lambda \ll 1/(4 \xi)$ we obtain $h_{*1} \simeq 1/\xi$ and $h_{*2} \simeq \lambda$. This scenario is not fully satisfactory since $\lambda$ itself is the effective CC at large scale factor values and, therefore, $\lambda$ cannot have its natural QFT value. For $\lambda<0$ there is no de Sitter behavior at large values of the scale factor. 
\item In the case when $w<-1$ we have $\xi<0$. The condition of reality of asymptotic values results in a requirement $\lambda>1/(4 \xi)$. We further have $h_{*1}-h_{*2}>0$, $h_{*2}<0$ and $h_{*1}>0$ for $\lambda>0$ whereas $h_{*1}<0$ for $1/(4 \xi)<\lambda<0$. The only case at which we could have a de Sitter behavior at large values of the scale factor is for $\lambda>0$. However, the asymptotic behavior is $\lim_{s \rightarrow 0} h = h_{*1}$ and at large $a$
there is no de Sitter behavior. 
\end{enumerate}

\section{The general conditions for a small effective CC and a hint from modified gravity}

\label{grav}

Many results of this paper, and especially the main results presented in subsection \ref{mainres}, stem from the analysis of the asymptotic behavior of the dynamics of $h$. The solutions for the function $h(s)$ which tend to a small positive constant value at large values of the scale factor have been interpreted as solutions of the cosmological constant in our approach. In this paper we use a specific form of the inhomogeneous EOS for the component $\rho$ with $\zeta(H)=\zeta_0 H^{\alpha}$. A natural and important following step would be to consider a broader class of functional behavior for $\zeta(H)$. In general the dynamics of the Hubble function would then be governed by the equation
\begin{equation}
\label{eq:dynH2gen}
s \frac{d h}{d s} + 3 (1+w) (h - \lambda - \xi' h^{1/2} \zeta(h))=0 \, ,
\end{equation}
where $\xi'$ is a constant.
A prerequisite for a solution of the CC problem for the function $\zeta(H)$ is the existence of a small and positive root $h_{*} \ll |\lambda|$ of the equation 
\begin{equation}
\label{eq:root}
h - \lambda - \xi' h^{1/2} \zeta(h)=0 \, .
\end{equation}
An additional condition is that the function $h$ asymptotes to $h_{*}$ at large values of the scale factor.

As already stated in section \ref{setup} and elaborated in \cite{Odin1}, a theory behind the inhomogeneous EOS of the type (\ref{eq:p}) could be some formulation of modified gravity. In the remainder of this section we perform an analysis of a  possible asymptotic behavior in a model of $f(R)$ modified gravity and discuss its implications for the solution of the CC problem.
For a review of the $f(R)$ modified gravity see \cite{Odinmodgrav} and \cite{Faraonimodgrav}.

We consider a $f(R)$ theory with a arbitrarily large cosmological constant energy density. In a universe with the FRW metric the dynamics of $H$ is given by the equation \cite{Faraonimodgrav}
\begin{equation}
\label{eq:modgravdyn}
3 f'(R) H^2 - \frac{1}{2}(R f'(R) - f(R))+3 H \dot{R} f''(R) = 8 \pi G \rho_{\Lambda} \equiv \Lambda \, ,
\end{equation}
where prime denotes the differentiation of $f(R)$ with respect to its argument and $R=12 H^2 + 6 \dot{H}$. Next we choose 
\begin{equation}
\label{eq:fR}
f(R)=R+b R^2-\frac{\mu^{2(n+1)}}{R^n} \, ,
\end{equation}
which satisfies the requirements of stability and positivity of the effective gravitational coupling \cite{Faraonimodgrav,Odinstab}. Here $b>0$, $\mu$ and $n$ are the parameters of the model. In general in this model, as well as in many other models of $f(R)$ gravity, there is not flat solution (with $R=0$). Furthermore, modified gravity theories can be subjected to stringent local gravity tests, e.g. measurements in the Solar system. We assume that the values of parameters $n$, $b$ and $\mu$ used here are consistent with the bounds from local gravity tests. Next we focus on the asymptotic behavior of $H$ and search for constant $H$ solutions of Eq. (\ref{eq:modgravdyn}). We neglect all time derivatives and  (\ref{eq:modgravdyn}) can now be written as 
\begin{equation}
\label{eq:1R}
3 H^2 - \Lambda - \frac{1}{2} \frac{\mu^{2(n+1)}}{(12 H^2)^{n}} \left( 1 + \frac{n}{2} \right) = 0 \, .
\end{equation}
Finally, for illustration purposes, we choose $n=1$ and a possible asymptotic value of $H$ is determined by the equation
\begin{equation}
\label{eq:n1}
H^4 - \frac{\Lambda}{3} H^2 - \frac{\mu^4}{48}=0 \, ,
\end{equation}
the solutions of which are
\begin{equation}
\label{eq:n1sol}
H^2_{*1,2}= \frac{1}{2} \left( \frac{\Lambda}{3} \pm \sqrt{\left( \frac{\Lambda}{3} \right)^2+\frac{\mu^4}{12}} \right) \, .
\end{equation}
For a negative $\Lambda$ with a large absolute value (so that $3 \mu^4/4 \Lambda^2 \ll 1$) the $+$ solution becomes 
\begin{equation}
\label{eq:nisolfin}
H^2_{*1} \simeq \frac{1}{16}\frac{ \mu^4}{ |\Lambda|} \, . 
\end{equation}
The comparison with the results of subsection \ref{mainres} shows a striking similarity with our model containing a component with inhomogeneous EOS. In the modified gravity model (\ref{eq:fR}) there is an asymptotic de Sitter behavior corresponding to a small $\Lambda_{eff}$. This finding strongly supports a conjecture that the relaxation of a large $|\Lambda|$ is also feasible directly in modified gravity theories. The details of the relaxation mechanism in $f(R)$ modified gravity theories will be elaborated elsewhere \cite{HS}. 


\section{Discussion}

\label{disc}

The preceding sections define our model, explain its inherent mechanism for the solution of the cosmological constant problem and outline its potential connection with the modified gravity theories. In this section we further discuss the aspects of the model which are of relevance for cosmological issues.

There are at least two sources of motivation for the  definition of component $\rho$ in terms of an inhomogeneous EOS. The first one is nonlinear (bulk) viscosity and the second one is modified gravity. Although both of these possibilities contain many specific variants, in this paper the focus has been on the approach valid for both of them.

As described in subsection \ref{mainres}, in suitable parameter regimes it is possible to have the solution of the CC problem for both positive and negative CC of large absolute value. Still, for negative $\Lambda$ the energy density $\rho$ remains positive throughout the evolution of the universe, whereas for positive $\Lambda$ the energy density $\rho$ must be negative. This is a strong signal that, at least for positive $\Lambda$, we should consider the component $\rho$ as an effective description of a more fundamental dynamics, possibly coming from the modifications of gravity. 

In our treatment we assumed that $\zeta_0>0$ in order to stay aligned with a possible interpretation that inhomogeneous term in (\ref{eq:p}) might come from some sort of nonlinear (bulk) viscosity. This assumption immediately relates the signs of parameters $1+w$ and $\xi$. With this approximation, the relaxation of a large positive CC requires $w<-1$, i.e. the parameter of the EOS should be of the phantom type \cite{Cald}. Since in modified gravity models it is possible to obtain the phantom-like effective dark energy \cite{Odinphant}, the interpretation of the component $\rho$ as an effective description of the modification of gravity gains further support. The requirement $\zeta_0>0$ allows us to simultaneously treat both sources of motivation for the inhomogeneous EOS of the component $\rho$. The relaxation of this requirement, i.e. allowing for negative values of $\zeta_0$ opens up space for scenarios that might be realized in specific theories of modified gravity. The study of models with negative $\zeta_0$ is an important challenge of future work. 

Quantum field theory provides various positive and negative contributions to $\Lambda$. Negative contributions come from the zero point energy of the fermionic degrees of freedom and some condensates. On the other hand, positive contributions come from the zero point energy of the bosonic degrees of freedom. In the analysis of the cosmological constant problem the size and interplay of positive and negative contributions is of considerable importance and so is the sign of the resulting $\Lambda$.  
An important conclusion of the present paper is that for any sign of $\Lambda$ in the model with the inhomogeneous EOS (\ref{eq:p}) we can end up in a universe with a small and positive effective cosmological constant. Whereas our model provides mechanism for the relaxation of both positive and negative CC, the interpretation of the component $\rho$ might differ.
The present analysis shows that for the relaxation of a negative $\Lambda$ with a large absolute value $\rho$ could equally play a role of a real cosmic fluid with nonlinear viscosity or be an effective description of the modified gravity effects. In the case of large positive $\Lambda$ the effective nature of $\rho$ seems more plausible.

The analysis of subsection \ref{mainres} further reveals that for both signs of the cosmological constant we can have two phases of accelerated expansion connected with an abrupt transition between them \footnote{For a negative $\Lambda$ we need to have $w<-1/3$ to have an accelerated expansion at small values of the scale factor.}. It is important to stress that these two phases of the accelerated expansion correspond to very different energy densities. It is an open question for future research whether these two phases of accelerated expansion and the abrupt transition between them could be useful in the description of the inflationary dynamics and the graceful exit.

From the analysis of the subsection \ref{mainres} we can see that the asymptotic value of the Hubble function $H$ depends on parameters $\xi$ and $\lambda$ i.e. on their ratio. Let us discuss our expectations of the size of these parameters. In the remainder of this paragraph we use the terms ``large" and ``small" loosely for illustration purposes. For the parameter $\xi$ we do not expect to be ``large". Namely, it describes either effects of nonlinear viscosity or deviations from general relativity. A natural size of $\xi$ could be described as ``small". On the other hand, from QFT we expect the absolute value of $\Lambda$ to be ``large". When $\xi$ is ``small" and $\lambda$ is ``large", the resulting value of $H^2$ is very small especially compared to $|\Lambda|/3$. Therefore, for proper values of $\alpha$ and for the parameters $\xi$ and $\lambda$ taking their ``natural" values we have a small asymptotic value of $H^2$. Within our model this solves the cosmological constant problem for any sign of $\Lambda$.

One of the most important issues towards a complete cosmological model which would incorporate the mechanism of the CC relaxation is the addition of matter and radiation components. The cosmological model should reproduce the eras of radiation domination and matter domination to be consistent with the available observational data. A dedicated analysis is required to account for details of the model with matter and radiation components. It will be particularly interesting to see how the addition of matter and radiation components interacts with the abrupt transition characteristic for the CC relaxation mechanism.
The cosmic coincidence problem could be possibly addressed only in such a full cosmological model. Although the full treatment is needed for the understanding of the entire dynamics of the full model, it is possible to argue that the addition of matter and radiation components will not affect the asymptotic behavior of the model and therefore the very CC relaxation mechanism. Namely, the energy densities of matter and radiation components decrease quickly with the expansion and at a sufficiently large scale factor value they become negligible. Then we are effectively back to the two component model presented in this paper and asymptotically we have the CC relaxation.

Finally, the focus of this paper is on the verification of the very effect of the relaxation of the large CC. There are many important questions that need to be addressed before our model with the inhomogeneous EOS could become a complete cosmological model. As already stated, it is important to learn if and how the behavior of the model changes when other cosmological components like radiation or nonrelativistic matter are added. Another question of considerable importance is which types of the inhomogeneous EOS apart from the one studied in this paper are capable of the relaxation of the large CC. The connection with the modified gravity theories seems especially worth pursuing. The growth and stability to perturbations is a relevant question too and so is the realization of the relaxation mechanism in astrophysical gravitationally bound systems. These questions are left for future work. 










\section{Summary and conclusions}

\label{conc}

The cosmological constant problem is a spot in theoretical physics landscape where the inadequacy of standard theoretical approaches is evident. An unconventional new ingredient is clearly called for. It is unclear, however, how big a deviation from the standard principles this new ingredient should represent. In this paper we have presented a simple approach based on a cosmological component with an inhomogeneous equation of state. The new ingredient is an inhomogeneous term in (\ref{eq:p}) which can be interpreted as nonlinear viscosity or the effect of modified gravity. In a particular parameter regime, which by itself requires no specific fine-tuning, the universe with a cosmological constant of any sign and an arbitrarily large absolute value asymptotically ends up in de Sitter regime with a small value of the effective cosmological constant. This result provides a solution of the CC problem without the need of fine-tuning. A preliminary analysis of a $f(R)$ modified gravity theory lends support to the claim that the mechanism of the CC relaxation studied for a model with EOS (\ref{eq:p}) also functions for modified gravity theories. The main results of this paper exemplify a scenario in which a large $\Lambda$ coming naturally from QFT calculations coexists with a small asymptotic value of the effective cosmological constant. The said results further open a possibility that the measured value of the cosmological constant is not the value coming from QFT, but it is determined by the QFT value. Apart from the intrinsic value of these results, they also allow us a bit different perspective on cosmological parameter puzzles: maybe instead of devising complex nonstandard ways of understanding the cosmological parameter values we should try to understand how the values of these cosmological parameters influence simple nonstandard dynamics. The relaxation mechanism for the cosmological constant presented in this paper hopefully follows the latter route.

{\bf Acknowledgements.} The author would like to thank N. Bili\'{c}, B. Guberina, R. Horvat and H. Nikoli\'c for useful comments on the manuscript. This work was
supported by the Ministry of Education, Science and Sports of the Republic of Croatia 
under the contract No. 098-0982930-2864.


\begin{thebibliography}{88}
\bibitem{SN} A.G. Riess et al., Astron. J. {\bf 116} (1998) 1009; S. Perlmutter et 
al., Astrophys. J. {\bf 517} (1999) 565;  W. Michael Wood-Vasey et al., Astrophys. J. {\bf 666} (2007) 694; Pierre Astier et al., Astron. Astrophys. {\bf 447} (2006) 31. 
\bibitem{WMAP}  E. Komatsu et al., arXiv:0803.0547 [astro-ph].
\bibitem{LSS} M. Tegmark et al., Astrophys. J. {\bf 606} (2004) 702; M. Tegmark et al., Phys. Rev. D {\bf 69} (2004) 103501.  
\bibitem{Rev} T. Padmanabhan, Phys. Rept. {\bf 380} (2003) 235; E.J. Copeland, M. Sami, S. Tsujikawa, Int. J. Mod. Phys. D {\bf  15} (2006) 1753; J. Frieman, M. Turner, D. Huterer, arXiv:0803.0982 [astro-ph]; T. Padmanabhan, arXiv:0807.2356 [gr-qc].
\bibitem{Wein} S. Weinberg, Rev. Mod. Phys. {\bf 61} (1989) 1.
\bibitem{Stra} N. Straumann, in Duplantier, B. (ed.) et al.: Vacuum energy, renormalization, 7-51, 
arXiv:astro-ph/0203330.
\bibitem{Nob} S. Nobbenhuis, arXiv:gr-qc/0609011.
\bibitem{Odin1} S. Nojiri, S.D. Odintsov, Phys. Rev. D {\bf 72} (2005) 023003.
\bibitem{Mota} D.F. Mota, C. van de Bruck, Astron. Astrophys. {\bf 421} (2004) 71.
\bibitem{Weinvisc} S. Weinberg, Astrophys. J. {\bf 168} (1971) 175.
\bibitem{Zim} W. Zimdahl, Phys. Rev. D {\bf 53} (1996) 5483.
\bibitem{Gron} \O. Gr\o n, Astrophys. Space Sci. {\bf 173} (1990) 191.
\bibitem{Odinmodgrav} S. Nojiri, S.D. Odintsov, Int. J. Geom. Meth. Mod. Phys. {\bf 4} (2007) 115.
\bibitem{Faraonimodgrav} T.P. Sotiriou, V. Faraoni, arXiv:0805.1726 [gr-qc]. 
\bibitem{Odinstab} S. Nojiri, S.D. Odintsov, Phys. Rev. D {\bf 68} (2003) 123512.
\bibitem{HS} H. \v Stefan\v ci\' c, in preparation.
\bibitem{Cald} R.R. Caldwell, Phys. Lett. B {\bf 545} (2002) 23.
\bibitem{Odinphant} F. Briscese, E. Elizalde, S. Nojiri, S.D. Odintsov, Phys. Lett. B {\bf 646} (2007) 105;
S. Jhingan, S. Nojiri, S.D. Odintsov, M. Sami, I Thongkool, S. Zerbini, Phys. Lett. B {\bf 663} (2008) 424. 
.
%


\end{thebibliography}
\end{document}